\documentclass[12pt,a4paper,final]{iopart}

\usepackage{iopams}  
\usepackage{graphicx}
\usepackage[breaklinks=true,colorlinks=true,linkcolor=blue,urlcolor=blue,citecolor=blue]{hyperref}

\def\beq{\begin{eqnarray}}
\def\eeq{\end{eqnarray}}
\def\beqa{\begin{eqnarray}}
\def\eeqa{\end{eqnarray}}

\begin{document}

\title[Pseudogap in cuprates driven by $d$-wave flux-phase order proximity effects]
     {Pseudogap in cuprates driven by $d$-wave flux-phase order proximity effects: A theoretical analysis from Raman and ARPES experiments}

\author{Andr\'es Greco and Mat\'{\i}as Bejas}
\address{Facultad de Ciencias Exactas, Ingenier\'{\i}a y Agrimensura and
Instituto de F\'{\i}sica Rosario (UNR-CONICET),
Avenida Pellegrini 250, 2000 Rosario, Argentina}
\eads{\mailto{agreco@fceia.unr.edu.ar} and 
      \mailto{bejas@ifir-conicet.gov.ar}}

\begin{abstract}
One of the puzzling characteristics of the pseudogap phase of high-$T_c$ cuprates is the nodal-antinodal dichotomy.
While the nodal quasiparticles have a Fermi liquid behaviour, the antinodal ones show non-Fermi liquid
features and an associated pseudogap.
Angle-resolved photoemission spectroscopy 
and electronic Raman scattering are two valuable tools which have shown 
universal features which are rather material-independent, and presumably intrinsic to the pseudogap 
phase.
The doping and temperature dependence of the Fermi arcs and the pseudogap observed by photoemission 
near the antinode 
correlates with the non-Fermi liquid behaviour observed by Raman for the B$_{1g}$ mode. 
In contrast, and similar to the nodal 
quasiparticles detected by photoemission, the Raman B$_{2g}$ mode shows Fermi liquid features. 
We show that these two experiments can be analysed, 
in the context of the $t$-$J$ model, by self-energy 
effects in the proximity to a $d$-wave flux-phase order instability. 
This approach supports a crossover origin for the pseudogap, and a scenario of two competing phases.
The B$_{2g}$ mode shows, in an underdoped case, a depletion at intermediate energy which has attracted a renewed 
interest. We study this depletion and discuss its origin and relation with the pseudogap.   
\end{abstract}




\submitto{\JPCM}

\section{Introduction} \label{sec:intro}

Most of the experimental probes in the normal state of underdoped (UD) cuprates
show signatures of a pseudogap (PG), a
gap-like feature that appears below a characteristic temperature
known generically as the pseudogap temperature $T^*$. Below this
temperature the electronic properties are very unusual \cite{timusk99}.
The value of $T^*$ may depend on the 
experimental probe, and it is far above the superconducting 
critical temperature $T_c$. 
In contrast to the behaviour of $T_c$, the PG and $T^*$ increase  
with decreasing doping in the UD region \cite{timusk99}.
Two main scenarios 
dispute the origin of the pseudogap. In one of them the PG can be associated 
with preformed pairs that exist above $T_c$. In the other one,  
the PG phase is distinct from superconductivity, and both compete \cite{hufner08,norman05}.
If the PG phase corresponds to a distinct order from superconductivity,
another issue is whether $T^*$ is a crossover or a
true thermodynamical transition temperature \cite{hufner08,norman05,tallon01}.
Among the experimental probes angle-resolved photoemission spectroscopy \cite{damascelli03,lee09} 
(ARPES) and electronic Raman scattering \cite{devereaux07} (ERS) 
have shown useful results for clarifying the nature of the nodal-antinodal dichotomy 
in the PG phase \cite{hufner08,norman05}.
Since these results are rather material-independent 
they can be assumed to be intrinsic features of the PG phase.
Thus, it is desirable that ARPES and ERS are described by the same theory.
In this paper we propose that the main features observed in ERS and ARPES can be analysed 
by self-energy effects in the proximity 
to the $d$-wave flux-phase instability \cite{affleck88,marston89,cappelluti99}. This
is the leading charge-ordered state that occurs in the framework of the 
$t$-$J$ model in the large-$N$ limit for the values of 
the electronic parameters suitable for cuprates \cite{bejas12}. 
Due to the fact that fluctuations of the flux-phase order parameter extend
above its  
critical temperature, in the present scenario the PG can be viewed as a precursor effect.
Thus,  no phase transition
occurs at $T^*$, and $T^*$ is a crossover temperature.

ARPES shows below a characteristic temperature, called here $T^{ARPES}$, 
a PG at the antinodal momentum of the Fermi surface ${\bf k}_F^{AN}$ that 
vanishes around the nodal ${\bf k}_F^N$.
Thus, instead of the expected large Fermi surface, disconnected Fermi arcs
are observed for which the length increases with increasing doping and 
temperature \cite{norman98,kanigel06,kanigel07,yoshida09,kondo09,hashimoto10,he11p}.
Using specific configurations of the incident and scattered electric field 
with respect to the crystallographic axis, the B$_{1g}$ and B$_{2g}$ symmetry modes can be individually 
selected by ERS. While the B$_{1g}$ mode probes electronic features around ${\bf k}_F^{AN}$, 
B$_{2g}$ probes the vicinity of ${\bf k}_F^N$ (see insets in figure \ref{fig:raman}) \cite{devereaux07}.
Similarly to ARPES, ERS shows unconventional features in the normal state in the UD region. 
The most salient features are: (a) 
With decreasing temperature the slope of the B$_{1g}$ response at $\omega=0$ 
first increases, and below a characteristic temperature, called here $T^{ERS}$, decreases. 
This characteristic temperature follows the same trend as $T^*$ \cite{sacuto13,venturini02,blumberg97}.
Since the slope of the Raman spectra at $\omega=0$ is proportional to the quasiparticle lifetime, 
the behaviour below $T^{ERS}$ is clearly non-Fermi liquid, showing the  
non-Fermi liquid nature of the quasiparticles (QPs) near ${\bf k}_F^{AN}$. 
(b) The B$_{2g}$ response shows a Fermi liquid behaviour, i.e., 
the slope increases with decreasing temperature showing the Fermi liquid nature of the 
QPs near ${\bf k}_F^N$ \cite{sacuto13,venturini02,blumberg97}.
(c) The B$_{2g}$ response shows a depletion at intermediate 
$\omega$ and relatively low doping \cite{sacuto13,nemetscheck97,opel00,gallais05}, and 
for which the origin and relation with the pseudogap has gained a renewed interest \cite{sakai13}.
We show here that the features observed by these two experiments can be analysed, 
in the context of the $t$-$J$ model, by self-energy 
effects in the proximity to the $d$-wave flux-phase instability. 

The paper is organized as follows. In section~\ref{sec:summary} we present a summary of the theoretical framework. 
In section~\ref{sec:comparison} we show the obtained results for ARPES and ERS and compare with the experiments.
In section ~\ref{sec:disconclu} we present the discussion and conclusion.

\section{Summary of the theoretical framework} \label{sec:summary}

In the present study we use the $t$-$J$ model 
\begin{equation}
H = -\sum_{i, j,\sigma} t_{i j}\tilde{c}^\dag_{i\sigma}
\tilde{c}_{j\sigma} + J \sum_{\langle i,j \rangle} \left( \vec{S}_i \cdot \vec{S}_j-\frac{1}{4} n_i n_j \right)\, 
\end{equation}
where $t_{i j} = t$ $(t')$ is the hopping between the first (second) nearest-neighbour 
sites on a square lattice, and  $J$ is the exchange interaction between the nearest-neighbour sites. 
$\langle i,j \rangle$ indicates a nearest-neighbour pair of sites. 
$\tilde{c}^\dag_{i\sigma}$ ($\tilde{c}_{i\sigma}$) is 
the creation (annihilation) operator of electrons 
with spin $\sigma$ ($\sigma = \downarrow$,$\uparrow$) 
in the Fock space without any double occupancy. 
$n_i=\sum_{\sigma} \tilde{c}^\dag_{i\sigma}\tilde{c}_{i\sigma}$ 
is the electron density operator and $\vec{S}_i$ is the spin operator.

In the framework of the path integral large-$N$ expansion \cite{foussats04,foussats06}, the $t$-$J$ model
shows, in leading order, a paramagnetic Fermi liquid phase with electronic dispersion
\begin{equation} \label{eq:ek}
\varepsilon_{\bf k}= -2 \left( t \frac{\delta}{2} + rJ \right) (\cos k_x+\cos k_y)-4 t'\frac{\delta}{2}\cos k_x \cos k_y - \mu,
\end{equation}
\noindent where $\delta$ is the hole doping away from half-filling, and 
$\mu$ is the chemical potential.
The chemical potential and $r$ are calculated self-consistently \cite{greco09,greco11} from
\begin{eqnarray}
r=\frac{1}{2 N_s} \sum_{\bf k} \cos(k_x) n_F(\varepsilon_{\bf k})
\end{eqnarray}
\noindent  and
\begin{eqnarray}
1-\delta=\frac{2}{N_s} \sum_{\bf k} n_F(\varepsilon_{{\bf k}})
\end{eqnarray}
\noindent where $n_F$ is the Fermi function and $N_s$ the number of sites.
The parameter $r$ contributes to the $J$-driven hopping term
$rJ$, which is the second term between parenthesis in (\ref{eq:ek}). This effective hopping
term,
which is not a bare electronic parameter, comes from the $J \sum_{\langle i,j \rangle} \vec{S}_i \cdot \vec{S}_j$
term of the $t$-$J$ model.
For more details about the path integral large-$N$ expansion see \cite{foussats04,foussats06}.

Hereafter, the appropriated parameters for cuprates $t'/t=-0.35$ and $J/t=0.3$ are used. 
The hopping $t$ and the lattice constant $a$ of the square lattice are the 
energy and length units, respectively.

The paramagnetic Fermi liquid phase is unstable against
a flux-phase \cite{affleck88,marston89,cappelluti99,bejas12}
below a transition temperature $T_{FP}$ which decreases
with increasing doping ending at a 
quantum critical point (QCP) [see solid line in figure \ref{fig:phasediag}(a)].
For $T<T_{FP}$, inside the flux-phase state, a true gap opens near $(\pi,0)$ and the new
Fermi surface consists of four small hole pockets near nodal direction.
The flux-phase can be considered, in the framework of the 
microscopic $t$-$J$ model, as a realization of the phenomenological proposed 
$d$-charge-density wave state \cite{chakravarty01}.
It is worth highlighting that, with increasing $T$ inside the flux-phase, the gap closes at $T_{FP}$ \cite{zeyher02}.
In addition, a pair breaking-like peak is predicted in the B$_{1g}$ mode in the normal state for 
$T<T_{FP}$ \cite{zeyher02,zeyher04}.
However, ARPES shows that the PG fills with temperature \cite{norman98,hashimoto10,he11p},
and ERS does not show clear signals of a peak in the normal state \cite{sacuto13,venturini02,blumberg97}.
We will discuss herein that a crossover origin for the PG could describe these features.

In the paramagnetic Fermi liquid phase and 
in the proximity to the flux-phase instability, the carriers are mainly 
dressed, in next-to-leading order [${O}(1/N)$], by the self-energy  
$\Sigma({\mathbf{k}},\omega)$ (see \cite{greco09,greco11}), where
\begin{eqnarray}\label{eq:SigmaIm0}
{\rm Im} \, \Sigma({\mathbf{k}},\omega)&=&-\frac{1}{N_{s}}\sum_{{\mathbf{q}}} \gamma^2({\bf q},{\bf {k}}) 
{\rm Im} \chi({\bf q},\omega-\varepsilon_{{\bf k-q}}) \nonumber \\
 && \hspace{2cm} \times \left[n_{F}(-\varepsilon_{{\bf k-q}}) + n_{B}(\omega-\varepsilon_{{\bf k-q}})\right], 
\end{eqnarray}
\noindent 
In (\ref{eq:SigmaIm0}) $n_B$  is the Bose factor, and 
\begin{equation}
\chi({\bf q},\omega)=[8 J r^2-\Pi({\bf q},\omega)]^{-1}
\end{equation}
\noindent  is the flux-phase susceptibility. 
$\chi({\bf q},\omega)$ diverges, at $\omega=0$ and ${\bf Q}=(\pi,\pi)$,  at $T_{FP}$. 
$\Pi({\bf q},\omega)$ is the electronic polarizability calculated with a form factor 
$\gamma({\bf q},{\bf k})=2 r J [\sin(k_x-q_x/2)-\sin(k_y-q_y/2)]$ \cite{greco09,greco11}.
Note that since the 
instability is  for ${\bf Q} = (\pi,\pi)$ 
the form factor $\gamma({\bf q},{\bf k})$ transforms 
into $\sim (\cos k_x -\cos k_y)$ which indicates the $d$-wave character of the flux-phase.

\section{Comparison with ARPES and ERS experiments} \label{sec:comparison}

\subsection{ARPES}

Using the Kramers-Kronig relation, $\mathrm{Re}
\Sigma({\bf k},\omega)$ can be determined  from $\mathrm{Im}
\Sigma({\bf k},\omega)$, and  the spectral function
$A({\bf k},\omega)$ computed as usual
\begin{eqnarray}\label{A}
A({\bf k},\omega)= -\frac{1}{\pi}\frac{{\rm Im}\Sigma({\bf k},\omega)}
{[\omega- \epsilon_{{\bf k}} - {\rm Re}\Sigma({\bf k},\omega)]^2 + [{\rm Im}\Sigma({\bf k},\omega)]^2}  
\end{eqnarray}

\begin{figure}
\centering
\includegraphics[width=8cm]{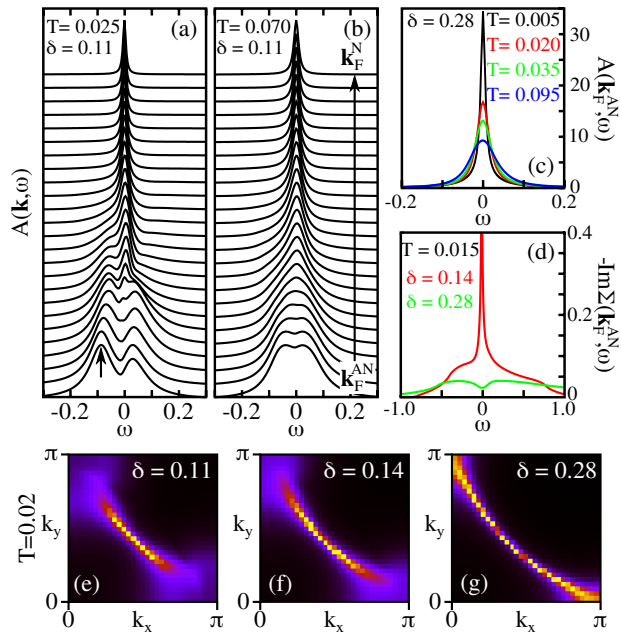}
\caption{
(a) and (b) Spectral functions from ${\bf k}_F^{AN}$ to ${\bf k}_F^{N}$ for doping 
$\delta=0.11$, and for $T=0.025$ and $T=0.07$, respectively. (c) Spectral
functions at ${\bf k}_F^{AN}$ for several temperatures for the overdoped case $\delta=0.28$. 
(d) $-{\rm Im} \Sigma$ at ${\bf k}_F^{AN}$ as a   
function of $\omega$ for $T=0.015$, and for $\delta=0.14$ and $\delta=0.28$. 
(e)-(g) Intensity plot for $A({\bf k},\omega=0)$ at $T=0.02$ for several dopings.   
}
\label{fig:arcs}
\end{figure}

In figures \ref{fig:arcs}(a) and \ref{fig:arcs}(b) we show $A({\bf k},\omega)$ along 
the Fermi surface, i.e., from ${\bf k}_F^{AN}$ to 
${\bf k}_F^N$,  
for the UD case $\delta=0.11$ at $T=0.025$ and $T=0.07$, respectively.
These calculations are in the paramagnetic Fermi liquid phase, where translational symmetry is not broken.
While the spectral functions 
near ${\bf k}_F^N$ are sharp and develops a well 
pronounced QP peak, approaching ${\bf k}_F^{AN}$ the spectral functions become broad. 
At low temperature ($T=0.025$) a PG is 
formed around ${\bf k}_F^{AN}$ [see arrow in figure \ref{fig:arcs}(a)]. With increasing 
temperature ($T=0.07$), the PG near 
${\bf k}_F^{AN}$ fills, as seen in the experiment \cite{kanigel06,kanigel07}.
Using the accepted 
value $t=400$ meV for cuprates the leading edge of the PG 
is $\sim 40$ meV, which is of the order of the experimental value.
In figure \ref{fig:arcs}(c) we show $A({\bf k},\omega)$ for several 
temperatures at ${\bf k}_F^{AN}$ for the overdoped 
(OD) case $\delta=0.28$.
In contrast to UD, the spectral functions do not show any signal of a PG and, as expected for a Fermi liquid, 
the intensity at $\omega=0$ decreases with increasing temperature. Figures \ref{fig:arcs}(e)-(g) show 
an intensity plot for 
$A({\bf k},\omega=0)$ for several dopings at $T=0.02$. 
While for OD the full Fermi surface is observed, 
for UD we obtain Fermi arcs. Clearly, the length of Fermi arcs 
increases with increasing temperature and doping. 
It is worth mentioning that the 
behaviour of $\Sigma({\bf k},\omega)$ near $\omega=0$ [figure \ref{fig:arcs}(d)] is the cause of the 
PG formation and the non-Fermi liquid behaviour of QPs near ${\bf k}_F^{AN}$.
While for overdoped $-{\rm Im} \Sigma$ shows the expected minimum at $\omega=0$ for both, 
${\bf k}_F^N$ (not shown) and ${\bf k}_F^{AN}$, for underdoped $-{\rm Im} \Sigma$ develops a pronounced 
maximum near ${\bf k}_F^{AN}$.

\begin{figure}
\centering
\includegraphics[width=8cm]{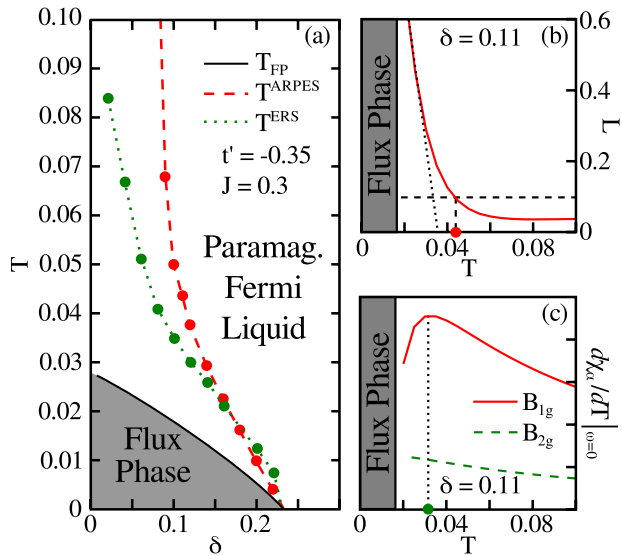}
\caption{
(a) Phase diagram in the $T$-$\delta$ plane. Dashed and dotted lines trace  
the crossover temperatures $T^{ARPES}$ and $T^{ERS}$, respectively. In the grey region below solid line ($T_{FP}$) 
the flux-phase is frozen. (b) The loss of the low energy spectral weight 
$L$ versus $T$ for $\delta=0.11$. The point at $T^{ARPES}=T \sim 0.04$ corresponds to the temperature
where $L=0.1$. Dotted line depicts the extrapolation of $L \rightarrow 0$. 
(c) Temperature dependence of the slope at $\omega=0$ for 
both, the B$_{1g}$ (solid line) and B$_{2g}$ (dashed line) modes. The point at $T^{ERS}=T \sim 0.035$
corresponds to the temperature where the slope of the B$_{1g}$ susceptibility at $\omega=0$ starts to decrease with decreasing $T$. 
}
\label{fig:phasediag}
\end{figure}

In figure \ref{fig:phasediag}(a), dashed line traces the temperature $T^{ARPES}$. 
For obtaining $T^{ARPES}$ we calculate the loss of the low energy spectral weight 
$L=1-I_0/I$ at ${\bf k}_F^{AN}$, 
where $I$ is the intensity 
of the spectral function at the leading edge of the PG [see arrow in figure \ref{fig:arcs}(a)], and $I_0$ 
the intensity at $\omega=0$ \cite{kanigel06,kanigel07}.
In figure \ref{fig:phasediag}(b) we plot $L$ versus $T$ for $\delta=0.11$, where $L$ tends smoothly 
to zero indicating a crossover behaviour instead of a true 
phase transition. For the case of a true phase transition $L$ should be exactly zero 
at the transition temperature.
We use the criterion that $T^{ARPES}$ is the temperature at which $L=0.1$.
In \cite{kanigel06,kanigel07} $T^{ARPES}$ was obtained 
extrapolating $L \rightarrow 0$. In our case this criterion will give a slightly different 
$T^{ARPES}$ [see doted line in figure \ref{fig:phasediag} (b)].
Note that $T^{ARPES} \sim 200$ K for $\delta=0.11$ is in a 
qualitative agreement 
with the experiments \cite{kanigel06,kanigel07}.
Finally, although it is a crossover 
temperature and not the onset of an abrupt transition, 
$T^{ARPES}$ terminates at the QCP. 

\begin{figure}
\centering
\includegraphics[width=8cm]{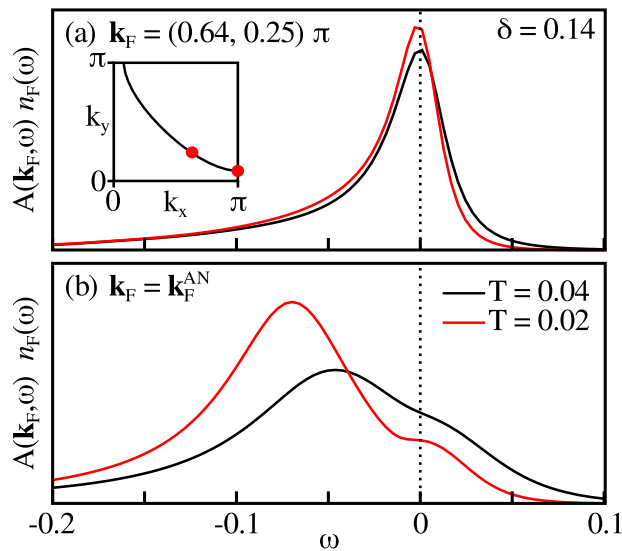}
\caption{$A({\bf k}, \omega) n_F(\omega)$ for (a)
${\bf k_F}=(0.64,0.25)\pi$ 
and  (b) ${\bf k}_F^{AN}$, for doping
$\delta=0.14$ and, for the two temperatures $T=0.02$
and $T=0.04$.
The inset in (a) shows the location of the
two momenta on the Fermi surface.
}
\label{fig:AknF}
\end{figure}

For comparison with the ARPES signal,
in figure \ref{fig:AknF} we show $A({\bf k}, \omega) n_F(\omega)$ for
${\bf k_F}=(0.64,0.25)\pi$ [panel (a)]
and  ${\bf k}_F^{AN}$ [panel (b)], for doping
$\delta=0.14$ and, for the two temperatures $T=0.02 < T^{ARPES}$
and $T=0.04 > T^{ARPES}$.
See also the inset of panel (a) for the location of the
two momenta on the Fermi surface.
Note that ${\bf k_F}=(0.64,0.25)\pi$ belongs to the arc region [figure \ref{fig:arcs}(f)]
for $T < T^{ARPES}$.
For ${\bf k_F}=(0.64,0.25)\pi$  we can see the expected behavior for a
well defined QP,
i.e., a well pronounced peak near $\omega=0$ and the fast decay for
$\omega > 0$ for both temperatures.
The flat experimental background seen in the experiment
was not included
along with any additional broadening due, for instance, to impurities.
The broadening of $A({\bf k},\omega)$ comes only from $\Sigma({\bf
k},\omega)$.
While the changes with temperature are small for ${\bf k_F}=(0.64,0.25)\pi$,
they are stronger for ${\bf k}_F^{AN}$. As mentioned above, the temperature
dependence
of $\Sigma({\bf k},\omega)$ at the antinode is stronger than at the node,
leading to
the PG for $T < T^{ARPES}$ and broad spectral functions.
The fact that the value for $A({\bf k}_F^{AN}, \omega) n_F(\omega)$ at $\omega=0$ is, in contrast to the
expected behaviour for a Fermi liquid, smaller for $T=0.02$
than for $T=0.04$ indicates the PG formation at low temperatures.

In summary, this scenario leads to a PG and Fermi arcs which depend on doping and temperature, as seen in the experiments. 
The PG occurs due to the coupling between QPs and the flux-phase soft mode 
in the proximity to the instability. Other works have proposed that antiferromagnetic \cite{sedrakyan10} and 
charge-density-wave \cite{perali96} 
fluctuations in the proximity
of the QCP may lead to 
a pseudogap.
In present approach the PG phase is
distinct from  superconductivity and associated with short-ranged and short-lived fluctuations in the proximity
to the flux-phase instability \cite{greco09,greco11}.
Since this picture occurs in the paramagnetic phase for $T>T_{FP}$ 
there is no breaking of the translational symmetry, as suggested by 
the experiments \cite{norman07}.
Finally, it should be noted that the self-energy (\ref{eq:SigmaIm0})
is not phenomenological.
This expression was derived beyond mean-field  
in a large-$N$ framework of a path integral representation for the $t$-$J$ model written 
in terms of Hubbard operators 
(see \cite{greco09} and \cite{greco11} and references therein). 
We suggest that the fact that our theory is supported by a controllable approach on the microscopic 
$t$-$J$ model is an advantage over phenomenological proposals.
 
\subsection{ERS}

The electronic Raman response is calculated by the following susceptibility 

\begin{eqnarray}\label{eq:ramansus}
\chi_{\alpha}(\omega)=\sum_{\bf k} \gamma_{\alpha}^2 \int d\varepsilon A({\bf k},\varepsilon) A({\bf k},\varepsilon-\omega) 
[n_F(\varepsilon-\omega)-n_F(\varepsilon)] 
\end{eqnarray}
\noindent where $\gamma_{\alpha}$ accounts for the Raman vertex in the geometry
$\alpha=$ (B$_{1g}$,B$_{2g}$) \cite{devereaux07}, i.e. 
$\gamma_{B_{1g}}= (t\delta/2+rJ) (\cos k_x-\cos k_y)$ 
and $\gamma_{B_{2g}}= 4 t'(\delta/2) \sin k_x  \sin k_y$.
In (\ref{eq:ramansus}) we have neglected vertex
corrections. A recent paper \cite{lin12} shows that vertex corrections are small at low energy 
where the PG opens. In addition, the consistency of our results with the experiments (see below) shows that 
(\ref{eq:ramansus}) captures important features.   

\begin{figure}
\centering
\includegraphics[width=8cm]{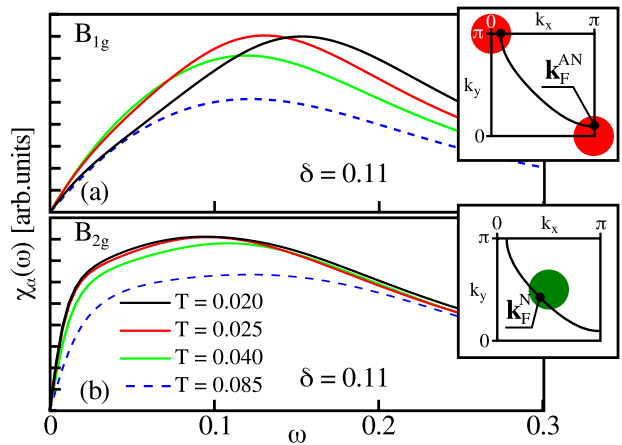}
\caption{
(a) and (b) B$_{1g}$ and B$_{2g}$ Raman response functions for $\delta=0.11$, respectively, for 
several temperatures. The inset in (a) shows  the location of ${\bf k}_F^{AN}$ 
and the region in the BZ selected by the B$_{1g}$ mode. The inset in (b) shows the same as (a) but for  ${\bf k}_F^{N}$ 
and the B$_{2g}$ mode. 
}
\label{fig:raman}
\end{figure}

Figures \ref{fig:raman}(a) and (b) show results for $\delta=0.11$ for the 
B$_{1g}$ and B$_{2g}$ Raman modes, respectively. 
For B$_{2g}$, which probes the QPs near ${\bf k}_F^N$, 
the results are consistent with the expected ones from a Fermi liquid, i.e., 
the slope at $\omega=0$ increases with decreasing $T$ [see dashed line in figure \ref{fig:phasediag}(c)]. 
For B$_{1g}$, which probes QPs near ${\bf k}_F^{AN}$, the situation is 
quite different. 
The slope first increases 
with decreasing $T$, and below a temperature identified here as $T^{ERS}$ 
this behaviour is reversed. The solid line 
in figure \ref{fig:phasediag}(c) shows the slope at $\omega=0$ of the B$_{1g}$ mode
versus $T$ for $\delta=0.11$. 
In figure \ref{fig:phasediag}(a), dotted line traces $T^{ERS}$ as a function of doping. 
Similar  
to $T^{ARPES}$, $T^{ERS}$ is a crossover temperature that terminates at the QCP.
In the OD side B$_{1g}$ and B$_{2g}$ show a Fermi liquid behaviour (not shown). 
It should be noted that high energy self-energy fluctuations, which may contribute to the flat 
spectrum
at high frequency \cite{sacuto13,venturini02,blumberg97}, are not considered. We focus on low energy features related to the pseudogap.

\begin{figure}
\centering
\includegraphics[width=8cm]{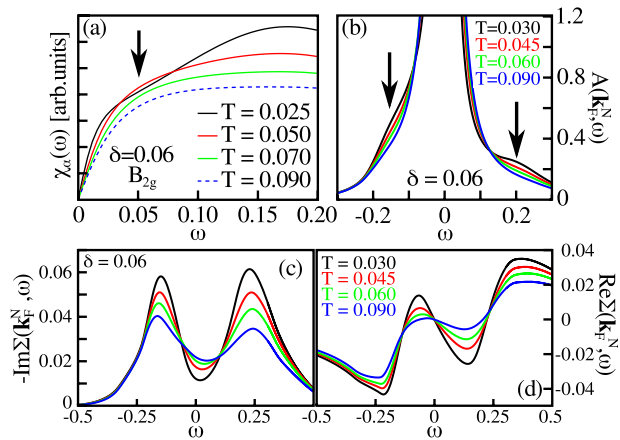}
\caption{
(a) B$_{2g}$ response function for the low doping $\delta=0.06$ and for several temperatures
showing a depletion at intermediate energy (arrow). 
(b) Spectral functions for several temperatures at ${\bf k}_F^N$, 
for $\delta=0.06$. Arrows show the side-band effects developed with decreasing temperature.
(c) and (d) $-{\rm Im} \Sigma$ and ${\rm Re} \Sigma$, respectively, at ${\bf k}_F^N$ for $\delta=0.06$ and several temperatures.
}
\label{fig:depletion}
\end{figure}

\begin{figure}
\centering
\includegraphics[width=8cm]{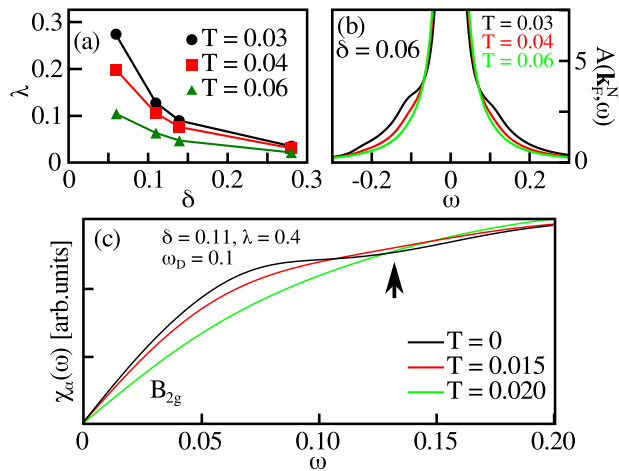}
\caption{
(a) The coupling $\lambda=- {\partial {\rm Re} \Sigma}/{\partial \omega}|_{\omega=0}$ as a function of $\delta$ for several
temperatures. (b) Considering a bosonic mode with the couplings $\lambda$ in (a) for $\delta=0.06$ the side-band features 
of figure \ref{fig:depletion} (b) can be qualitatively reproduced. 
(c) B$_{2g}$ response function for several temperatures
obtained assuming a bosonic mode with $\lambda=0.4$ and
$\omega_D=0.1$ for $\delta=0.11$.
}
\label{fig:bosmode}
\end{figure}

As mentioned in the introduction, ERS shows  
a depletion \cite{sacuto13,nemetscheck97,opel00,gallais05} at intermediate energy for $B_{2g}$ which
has recently attracted a renewed interest \cite{sakai13}.
In \cite{sakai13} the origin of this depletion was discussed in terms of a 
$s$-wave pseudogap. In the following we will propose an alternative explanation for the 
B$_{2g}$ depletion.
Figure \ref{fig:depletion}(a) shows results for the $B_{2g}$ mode for the low doping $\delta=0.06$. 
Besides the increasing of the slope at $\omega=0$ with decreasing
$T$, a depletion at intermediate energy is obtained (see arrow) below $T \sim 0.05 \sim 230$ K. 
Since B$_{2g}$ probes the nodal QPs, we identify this effect with the behaviour of $A({\bf k},\omega)$ 
near ${\bf k}_F^N$. 
In figure \ref{fig:depletion}(b) we show results for the spectral function at ${\bf k}_F^N$ for 
$\delta=0.06$ and several temperatures. The spectral 
functions show 
well defined QPs and, additionally, with decreasing $T$ side-bands are developed 
(see arrows). These self-energy side-band effects, 
which reduce the QP weight at ${\bf k}_F^N$, are 
responsible for the depletion in B$_{2g}$ at intermediate energy.  
Similar to the spectral function, $\Sigma({\bf k},\omega)$ at ${\bf k}_F^N$ [figures \ref{fig:raman}(c) and (d)] is Fermi liquid-like :
$-{\rm Im}\Sigma({\bf k}_F^N,\omega)$ shows always a minimum at $\omega = 0$ and some structure at $\omega \; \pm \sim 0.2$, 
and ${\rm Re}\Sigma({\bf k}_F^N,\omega)$ shows a negative slope at $\omega = 0$.
Thus, it is possible to extract the effective mass or dimensionless coupling parameter 
$\lambda=-{\partial {\rm Re} \Sigma}/{\partial \omega}|_{\omega=0}$. Since the Fermi liquid negative slope of ${\rm Re} \Sigma$ at $\omega=0$ increases with 
decreasing temperature, $\lambda$ increases with decreasing $T$.  
In figure \ref{fig:bosmode}(a) we show 
$\lambda$ versus doping for several temperatures. 
Using the values for $\lambda$ extracted for $\delta=0.06$ the side-band features of figure \ref{fig:depletion}(b) 
can be qualitatively reproduced [figure \ref{fig:bosmode}(b)].  

Figure \ref{fig:bosmode}(a) shows that $\lambda$ 
is strongly doping  and temperature dependent, making the depletion at intermediate energy hardly visible for 
large doping.
See for instance figure \ref{fig:raman}(b) for $\delta=0.11$ where the depletion at intermediate energy 
is not visible. 
This doping dependence of $\lambda$ is consistent with the fact that the depletion is not observed experimentally 
for large doping \cite{sacuto13,nemetscheck97,opel00,gallais05}.
Thus, the flux-mode which is very dominant at 
${\bf q} \sim (\pi,\pi)$ and leads
to the pseudogap near the antinode, possesses residual Fermi liquid-like interactions near the node.
However, these residual interactions vanish with doping faster than the pseudogap.
As in \cite{sakai13}, our mechanism for the depletion has an origin on 
electronic correlation effects. It is linked to the PG in the sense that both, the depletion 
and the PG,  
come from the same self-energy.
However, it would be precipitous to identify the B$_{2g}$ depletion as the key to understanding the PG. 
For instance, the B$_{2g}$ depletion disappears, in the experiment and in present theory, much faster with doping 
than the PG at ${\bf k}_F^{AN}$. We note that in \cite{sakai13} the B$_{2g}$ depletion 
was studied only for doping $\delta=0.05$. It will be interesting to discover
whether the features obtained in \cite{sakai13} remain robust with increasing doping. 

In spite of the qualitative agreement between present theory and the
experiments on the B$_{2g}$ depletion at intermediate energy, 
there are some unresolved issues arising from the phenomenology. Another possible
origin for the depletion lies on the coupling between QPs and a 
bosonic mode \cite{gallais06}.
In \cite{gallais06} a doping dependence of 
the bosonic coupling parameter $\lambda$ was phenomenologically proposed in order to describe 
the experiments. It is worth mentioning that the proposed $\lambda$ versus doping of 
\cite{gallais06} is close to the values 
obtained by ourselves [figure \ref{fig:bosmode}(a)].
In figure \ref{fig:bosmode}(c), using
the band dispersion $\varepsilon_{\bf k}$ for $\delta=0.11$, 
we show the B$_{2g}$ depletion (see arrow) caused by a bosonic mode
of characteristic frequency $\omega_D=0.1$ and $\lambda=0.4$.
Since B$_{2g}$ probes the nodal direction, the depletion could 
have the same origin as the nodal kink observed in ARPES \cite{zhou05,park13,park14}.
Although it 
is controversial whether the origin of the kink is
magnetic \cite{manske01} or phononic \cite{lanzara01,zeyher01}, it is fairly 
accepted that one possible
origin is due to the coupling between QPs and a bosonic mode.
Since the ARPES kink is observed from underdoped to overdoped \cite{park14},
the question is why the bosonic mode associated with the kink is not seen in B$_{2g}$
for the same doping range.
On the other hand, if the bosonic mode is phononic in origin
and isotropic in ${\bf k}$-space as expected from phonons \cite{zeyher01}, a
similar depletion
should also be observed in B$_{1g}$, at least in overdoped where 
the low energy spectrum is not affected by the PG, and shows a Fermi liquid behaviour as B$_{2g}$.  
We conclude that more experiments and theory might be focused to discover why the excitations involved in the nodal 
kink are not clearly seen in ERS in the full range of doping.

\section{Discussion and conclusion} \label{sec:disconclu}

In this paper we have discussed the PG as a precursor effect approaching the
flux-phase instability. In this scenario $T^*$ can be viewed as a crossover
temperature and not as a true thermodynamical transition temperature, and the PG originates
due to
fluctuations which extend above the critical temperature $T_{FP}$.
Thus, the PG occurs without the breaking of the translational symmetry.
Nevertheless,
it is important to bear in mind that present scenario requires that at
sufficiently
low temperature the flux-phase becomes thermodynamically stable below
$T_{FP}$. Below $T_{FP}$ the translational symmetry is broken and a 
Bragg reflection at $(\pi,\pi)$ is expected, which is controversial from the experimental point of view.
Thus, in order that our scenario be more reliable in comparison with experiments
one
should expect that the $T_{FP}$ line exists at very low temperature and is sunk
below the superconducting dome, being the normal state only affected by fluctuations. 
In addition, we should mention that below $T_{FP}$ the Fermi surface transforms to
closed Fermi pockets \cite{greco09} for which observation is also controversial from ARPES.
It should be noted that the reported Fermi pockets
observed in quantum oscillations experiments at low temperature
were discussed in the context of the $d$-charge-density-wave picture \cite{chakravarty08} 
which shows similar pockets to
those expected from present theory below $T_{FP}$.

In conclusion, we have shown that self-energy effects  
in the proximity of a $d$-wave flux-phase instability can account for the main features observed in ARPES and 
ERS experiments. It was shown that the doping and temperature dependence of the Fermi arcs and 
the pseudogap at the antinode
correlate with the behaviour of the B$_{1g}$ mode. 
In addition, the behaviour of 
the quasiparticles at the node correlates with the B$_{2g}$ mode.  
We have also discussed the depletion at intermediate energy observed in the B$_{2g}$ spectra.

\ack
The authors would like to thank M Le Tacon, A Muramatsu, H Parent, and H Yamase for their useful discussions.

\section*{References}
\bibliography{mainJPCM.bib}

\end{document}